\documentclass{workshop}
\usepackage{nohyperref}

\begin{document}

\title{
GRB Prompt Emission Spectra: The Synchrotron Revenge
}

\author{
Maria Edvige Ravasio$^1$ $^2$
\\[12pt]  
%
$^1$  Universit\`a degli Studi di Milano-Bicocca, Dipartimento di Fisica U2, Piazza della Scienza, 3, I–20126, Milano, Italy\\
$^2$  INAF – Osservatorio Astronomico di Brera, via Bianchi 46, I–23807 Merate (LC), Italy \\
%
\textit{E-mail: m.ravasio5@campus.unimib.it, maria.ravasio@inaf.it}
}

\abst{
 After more than 40 years from their discovery, the long-lasting tension between predictions and observations of GRBs prompt emission spectra starts to be solved. We found that the observed spectra can be produced by the synchrotron process, if the emitting particles do not completely cool. Evidence for incomplete cooling was recently found in Swift GRBs spectra with prompt observations down to 0.5 keV (Oganesyan et al. 2017, 2018), characterized by an additional low-energy break. 
In order to search for this break at higher energies, we analysed the 10 long and 10 short brightest GRBs detected by the Fermi satellite in over 10 years of activity. We found that in 8/10 long GRBs there is compelling evidence of a low energy break (below the peak energy) and the photon indices below and above that break are remarkably consistent with the values predicted by the synchrotron spectrum (-2/3 and -3/2, respectively). None of the ten short GRBs analysed shows a break, but the low energy spectral slope is consistent with -2/3. Within the framework of the GRB standard model, these results imply a very low magnetic field in the emission region, at odds with expectations. 
I also present the spectral evolution of GRB 190114C, the first GRB detected with high significance by the MAGIC Telescopes, which shows the compresence (in the keV-MeV energy range) of the prompt and of the afterglow emission, the latter rising and dominating the high energy part of the spectral energy range. 
}

\kword{workshop: proceedings --- prompt emission -- gamma-ray burst -- radiation mechanism}

\maketitle
\thispagestyle{empty}

\section{Introduction}
The nature of the radiation mechanism behind the prompt emission of Gamma-Ray Bursts (GRBs) has puzzled astronomers since their discovery and still remains not fully understood after more than 40 years of observations. 
The non-thermal shape of the observed spectra and the likely presence of accelerated particles in a magnetized region lead to the suggestion that the synchrotron process could be the responsible for such $\gamma$-ray emission \citep{ReesMeszaros1994, Katz1994, Tavani1996, SariPiran1997}. However, the inconsistency between the observed GRB spectra 
and the theoretical spectral shapes expected for synchrotron emission has kept the debate alive for many years.

From an observational point of view, the typical GRB prompt emission spectrum is usually fitted with the so-called Band function \citep{Band1993}, namely two power-laws with slopes $\alpha$ and $\beta$, smoothly connected at the peak energy $E_{peak}$ of the $\nu F\nu$ spectrum. The slope $\alpha$ of the low-energy power-law has been found to have an average value of $\langle \alpha \rangle \sim$ --1 for long bursts, while 
short GRBs, on average, are harder: 
$\langle \alpha \rangle \sim$ --0.4 \citep{Preece2000, Kaneko2006, Ghirlanda2009, Nava2011, Goldstein2012, Gruber2014}. 

For the typical physical parameters expected in the prompt GRB emitting region, the electrons should radiate in a regime of fast cooling \citep{Ghisellini2000} and the observed spectrum should have a photon slope $\alpha=-3/2$ below the peak energy $E_{peak}$.
Thus, most of the typical observed slopes are inconsistent with the synchrotron predictions, since 
the majority of GRBs are
harder than --3/2, and in some case even harder than --2/3, corresponding to the limiting case of the single particle spectrum, the so-called synchrotron 'line-of-death' \citep{Preece1998}.
The observed hard value of the low-energy photon index is one of the key observational features that strongly challenged 
the synchrotron interpretation.

Recently, extending the investigation of the prompt emission of 34 long GRBs down to the soft X-rays, \citet{Oganesyan2017} found low-energy breaks in their spectra. The distribution of these breaks is centered around $E_{break} \sim$ 30 keV and the slopes below and above that break are distributed around $\langle \alpha_{1} \rangle$= --0.51 ($\sigma=$0.24) and $\langle \alpha_{2} \rangle$=--1.56 ($\sigma$=0.26). 
These slopes are consistent, within 1$\sigma$, with the expectations of the synchrotron theory ($\alpha_{1}$= --2/3 and $\alpha_{2}$= --3/2) in a marginally fast cooling regime (i.e., when $\nu_{cool} \sim \nu_{min}$ \citep{Daigne2011, Beniamini2013}).
Since this low-energy break was found in the soft X-rays, we decided to extend the search of 
this feature also at higher energies through GRBs detected by the Fermi/GBM instrument (8 keV - 40 MeV).

We analyzed GRB 160625B \citep{Ravasio2018}, a very bright burst whose light curve is composed by three distinct emission episodes: a precursor, the main event and a last dimmer event. 
We performed a time-resolved spectral analysis on the main event 
using an empirical function composed of three power-laws smoothly connected at two breaks. 
Fitting with three power-laws actually provides an improvement of the fit (with respect to a function with only two power-laws) in the majority of the spectra analysed ($\sigma \ge $ 8, according to the F-test).
The time-resolved spectra of this burst are characterised by a low-energy power-law photon index centered around $\alpha_1=-0.63$ ($\sigma$=0.08), the presence of an additional low-energy break at $E_{break} \sim$ 50-100 keV, a second power-law photon index centered around $\alpha_2=-1.48$ ($\sigma$=0.09) between the break and the peak energy, a second spectral break (representing the peak in $\nu F_{\nu}$) varying with time in the range $E_{peak} \sim$ 300 keV -- 6 MeV, and a third power-law photon index $\beta \sim-2.6$ describing the spectrum above the peak energy. The slopes below and above the break are remarkably consistent with the synchrotron predicted values.

Motivated by these results, we performed a systematic search for this feature in a larger sample of GRBs \citep{Ravasio2019}. To this aim, we selected the 10 brightest long GRBs and the 10 brightest short GRBs detected by the GBM instrument over 10 years of activity. 

\begin{figure}[t]
\centering
\includegraphics[width=9cm]{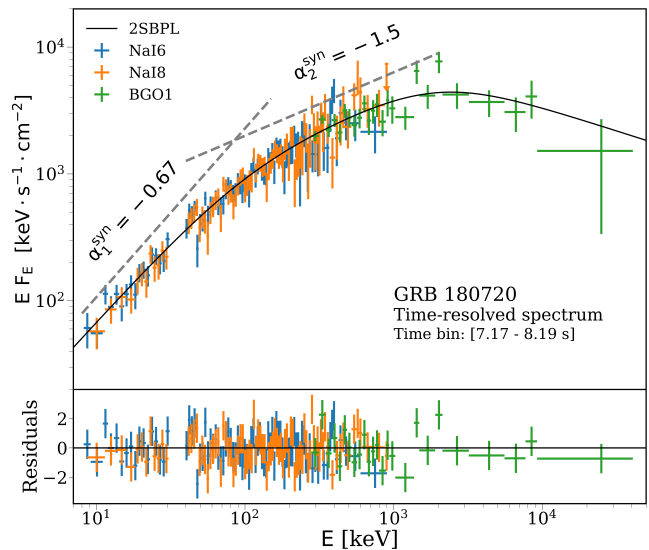}
\caption{Example of a spectrum best fitted by a 2SBPL (i.e. three power laws  smoothly  connected  at  two  breaks). The best fit values of the 2SBPL model parameters $\alpha_1$= --0.71, $E_{break}$=93.6 keV, $\alpha_2$= --1.47,$E_{peak}$=2.4 MeV, and $\beta$=-2.38. The two dashed lines show the power laws with the photon indices predicted by synchrotron emission. Data-to-model residuals are shown in the bottom panel. From \citet{Ravasio2019}.}
\label{fig:180720Bspectrum}
\end{figure}

For long GRBs, the time-integrated analysis shows that in 8/10 of the brightest GBM bursts,  the standard 
fitting function fails to provide an acceptable fit: data require an additional spectral break $E_{break}$, located between $\sim$ 10 keV and 300 keV. For these eight GRBs, a detailed time-resolved analysis revealed that $\sim$ 70\% of the spectra analysed also show strong evidence of an additional low-energy spectral break. The photon index of the power law below $E_{break}$ is distributed around $\langle \alpha_1 \rangle$=--0.58 ($\sigma$=0.16) while the power law between $E_{break}$ and $E_{peak}$ has photon index $\langle \alpha_2 \rangle$=--1.52 ($\sigma$=0.20), both remarkably consistent with the predicted values for synchrotron emission (see Fig. \ref{fig:180720Bspectrum} for an example of a typical spectrum). 

\begin{figure}[t]
\centering
\includegraphics[width=9cm, height=5cm]{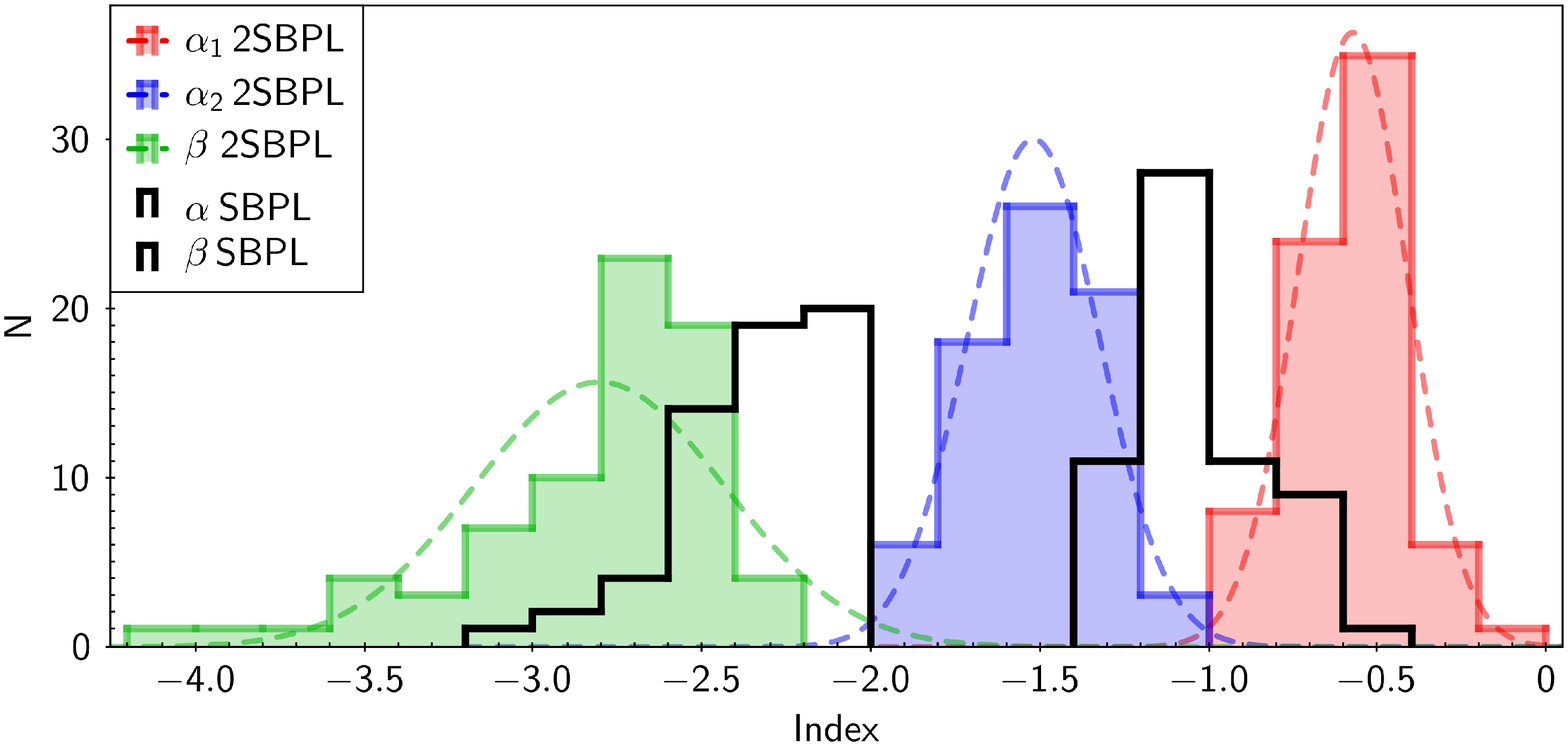}
\caption{Distribution of the spectral indices for the time-resolved fits of the 8 long GRBs showing an additional spectral break. The spectral indices $\alpha_1$, $\alpha_2$ and $\beta$ of the 2SBPL model are shown with red, blue, and green filled histograms, respectively. Gaussian functions showing the central value and standard deviation of the distributions are overlapped to the histograms (colour-coded dashed curves). The black empty histograms represent the distributions of the two photon indices $\alpha$ and $\beta$ of the SBPL model, for spectra where the SBPL is the best fit model. From \citet{Ravasio2019}.}
\label{fig:histo_ind_long}
\end{figure}

The remaining time-resolved spectra ($\sim$30\%) are best fitted by the standard 
function. In these cases, one power law is sufficient to model the spectra below $E_{peak}$, and the mean value of its photon index is $\langle \alpha \rangle$=--1.02 ($\sigma$=0.19). 
This value is in agreement with previous works and, interestingly, it lies between the values of $\alpha_1$ and $\alpha_2$, as shown in Fig. \ref{fig:histo_ind_long} by the black empty histogram. 
Speculating that most of the spectra show a break below the peak energy, the value of $\alpha$ can be understood as a sort of average value between $\alpha_1$ and $\alpha_2$.
The consistency with synchrotron emission in the long GRBs spectra is supported also by the recent works of \citet{Oganesyan2019} and \citet{Burgess2019}, where a physical synchrotron model has been found to fit well the spectra. Note, however, that the majority of spectra analysed by \citet{Burgess2019} displays a regime of slow cooling, implying a even lower efficiency of radiation.

Regarding short GRBs, 
our analysis shows that no 
spectra 
shows a second break at low energies. They are best-fitted by the standard fitting function, namely with only one component below the peak energy. The low-energy slope of this component is centered around $\langle \alpha \rangle=-0.78$ ($\sigma$=0.23). As for $\alpha_1$ in long GRBs, this value is consistent with the low-energy synchrotron photon index --2/3.

These results suggest that the underlying population of electrons does not cool completely. In fact, let us assume a power law distribution of 
electrons, injected above some injection energy $\gamma_{min}$: they cool via synchrotron process (and the observed slope --3/2 below $E_{peak}$ is the footprint of their cooling), but then at some energy $\gamma_{cool}$ they stop cooling, and the 
single electron spectral slope  --2/3 appears. 
Therefore, interpreting $E_{break}$ as the synchrotron cooling frequency $\nu_{cool}$, the implied magnetic field $B^\prime$ in the (comoving) emitting region is small, i.e. between 1 and 40 G.
The situation is even more extreme in short GRBs. 
Contrary to long GRBs, their spectra are described by only one power law below 
$E_{peak}$, with slope $\sim$ --2/3,
suggesting that 
$\nu_{cool}\simeq \nu_{min}$ and the cooling is even more incomplete, thus yielding an even smaller estimate of $B^\prime$.

In the framework of the GRB standard model, the expected magnetic field, in the emitting region, should be of the order of $10^{6}$ G, i.e. much larger than the estimates above. If the emitting region were at a standard distance (R $\sim 10^{13}$ cm), 
a small 
$B\prime$ would imply a strong inverse Compton component, whose luminosity would exceed the 
synchrotron one by a factor of $\sim 10^{7}$. 
A weak magnetic 
field could be possible if the emitting region were at larger distances (i.e.  $10^{18}$ cm) where, however, the afterglow emission could also take place , and the expected variability timescale would be much larger than the typical short values observed in the prompt emission. 

Recently, \citet{Ghisellini2019} proposed  that spectral results of \citet{Oganesyan2017}, \citet{Oganesyan2018}, \citet{Ravasio2018} and \citet{Ravasio2019} 
could be explained considering synchrotron emission from  protons rather than electrons. 
Due to their larger mass, the synchrotron cooling timescale of protons can be relatively long, i.e. $\sim 10^8$ times longer than electrons. This can account for $\nu_{cool} \sim$ 100 keV with a standard magnetic field of $B^\prime \sim 10^6$ G and considering the emission region at R $\sim 10^{13}$ cm (thus accounting for a short timescale variability).

\section{The $\gamma$-ray extra-component in GRB 190114C}

GRB 190114C is the first GRB detected with high significance at TeV energies by the MAGIC telescopes \citep{Mirzoyan2019}. 
It was also detected by Fermi/GBM with a [10-1000 keV] fluence $f=4.433 \pm 0.005 \times10^{-4}$ erg/c$m^2$.
We performed  \citep{Ravasio2019b} a detailed time-resolved spectral analysis of the Fermi/GBM emission, from 10 keV to 40 MeV,
up to $\sim$60 s from the trigger time.

We found that the first 4.8 seconds 
are characterized by the standard prompt spectral shape, fitted by a standard 
function with typical parameters. After 4.8 seconds from the trigger, 
there is evidence of the emergence of another spectral component, superimposed on the typical prompt shape spectrum. This component rises and decays quickly, reaching its peak flux 
($F=1.7\times10^{-5} erg/cm^2/s$, in the 10 keV-40 MeV energy range) at $\sim$ 6 s. 
We found that this additional component is well fitted by a power law of spectral index $\Gamma_{PL} \sim -2$ (see Fig. 1 in \citet{Ravasio2019b}).

We interpreted this component as the afterglow of the burst, because i) it appears after the trigger of the prompt event, and peaks when most of the prompt emission energy has already been radiated, ii) it lasts much longer than the prompt emission, iii) it is characterized by a spectral index ($\Gamma_{PL}\sim$-2) typical of known afterglows, iv) with the exception of the early variable phases, its light curve smoothly decays 
as $F \propto t^{-1}$, typical of the known afterglows.

Therefore, we found the evidence of the co-presence in the keV-MeV energy range of both the prompt and the afterglow components, already after $\sim$ 5 s from trigger time. We interpret the peak of the afterglow as the fireball deceleration time 
and estimate the bulk Lorentz factor during the coasting phase $\Gamma_0$=700 or 130, (assuming an homogeneous or wind medium, respectively).

\label{last}

\end{document}